\documentstyle[epsf]{mn}           
\def \th {\thinspace}
\def \sun {\hbox {$\odot$}}
\def \src {\hbox {GRO\thinspace J1655-40}}

\def \degmark {^\circ}

\def \hcm {\hbox {\ifmmode $ atoms cm$^{-2}\else atoms cm$^{-2}$\fi}}

\def\approxgt{\mathrel{\hbox{\rlap{\lower.55ex \hbox {$\sim$}}
        \kern-.3em \raise.4ex \hbox{$>$}}}}
\def\approxlt{\mathrel{\hbox{\rlap{\lower.55ex \hbox {$\sim$}}
        \kern-.3em \raise.4ex \hbox{$<$}}}}

\begin{document}

   \title[Discovery of a disk line in GRO\th J1655-40]             
         {Discovery of a red and blue shifted iron disk line \\
          in the galactic jet source GRO\th J1655-40}

   \author[M. Ba\l uci\'nska-Church et al.]    
      {M. Ba\l uci\'nska-Church and M. J. Church\\
      University of Birmingham, School of Physics and Astronomy,
      Birmingham, B15 2TT, UK\\
      e-mail: mjc@star.sr.bham.ac.uk, mbc@star.sr.bham.ac.uk\\}

\date{Accepted 17 December 1999. Received 24 August, 1999}
\maketitle

\begin{abstract}
       We report the discovery of emission features in the X-ray
spectrum of \src\ obtained with {\it RXTE} during the observation of
1997, Feb 26. We have fitted the features firstly by two Gaussian lines 
which in four spectra analysed have average energies of $\rm {5.85\pm 0.08}$ keV and 
$\rm {7.32\pm 0.13}$ keV, strongly suggestive that these are the red
and blue shifted wings of an iron disk line. These energies imply a
velocity of $\sim $0.33 c. The blue wing is less
bright than in the calculated profiles of disk lines near a black hole
subject to Doppler boosting, however known Fe absorption lines in
\src\ at energies between $\sim $7 and 8 keV can reduce the apparent 
brightness of the blue wing. Secondly, we have fitted the spectra
using the disk line model of Laor based on a full relativistic treatment
plus an absorption line, and show that good fits are obtained. This 
gives a restframe energy of the disk line between 6.4 and 6.8 keV 
indicating that the line is iron $\rm {K_{\alpha}}$ emission probably of 
significantly ionized material. The Laor model shows that the line
originates in a region of the accretion disk extending from $\sim $10
Schwarzschild radii from the black hole outwards. The line is direct evidence for the black hole
nature of the compact object and is the first discovery of a highly 
red and blue shifted iron disk line in a Galactic source.

\end{abstract}


\section{Introduction}
The X-ray nova \src\ discovered with BATSE (Zhang et al. 1994) is one
of the two only definite Galactic superluminal jet sources: \src\ and 
GRS\thinspace 1915+105 (Mirabel \& Rodriguez 1994). These two sources 
are also transient whereas the source SS\thinspace 433 with velocity in the jets of 0.26 c
is persistent. Radio images of the source showed condensations 
moving in opposite directions (Tingay et al. 1995). The apparent superluminal motion 
implies that the emitting plasma has velocity close to c, in fact 0.92 c
(Hjellming \& Rupen 1995). Initially, X-ray outbursts were
separated by $\sim $120 d (Zhang et al. 1995), and 
on 1996, April 25 an outburst began lasting 16 months as 
shown by the {\it RXTE} ASM (Sobczak et al. 1999). By the time of this
outburst, the strong radio activity had ceased, although radio
emission was detected again on 1996, May 29 (Hunsted \&
Campbell-Wilson 1996). Lack of radio detection after that date despite
regular monitoring (Tingay 1999) implies that the jets had ceased to exist.

The optical observation of 1996, March provided a mass of the central objects 
of $\rm {7.02\pm 0.22}$ M$_{\sun}$ and an inclination angle of 
$\rm {69.5\pm0.08^{\degmark}}$ (Orosz \& Bailyn 1997). An inclination of
$\rm {67.2\pm3.5^{\degmark}}$ was found by van der Hooft et al. (1998). \src\ is 
thus generally accepted to be a Black Hole Binary, the only Galactic jet source 
with mass determination evidence for its black hole nature.

Several observations were made with {\it ASCA}, the first on 1994,
August 23 and on 1997, Feb 26--28 an observation lasting one orbital
cycle was made during which the {\it RXTE} observation discussed here
was also made. Iron absorption line features were found at $\sim $6.6 keV and
$\sim $7.7 keV when the source was less bright (0.27--0.57 crab), and
at $\sim $7 keV and $\sim $8 keV when the source was brighter (2.2
crab), and these were identified with $\rm {K_{\alpha}}$ and $\rm
{K_{\beta}}$ lines of Helium-like and Hydrogen-like iron respectively.
In the observation of 1997, Feb 26--28, a broad absorption feature at
$\sim $6.8 keV was seen, thought to be a blend of He-like and H-like
lines (Yamaoka et al. 1999). Sobczak et al. (1999) found evidence for 
an iron edge at 8 keV. X-ray dipping has been observed in \src\, and many
short and deep dips similar to the those in \hbox{Cygnus\th X-1}
(Ba\l uci\'nska-Church et al. 1999) are seen. 
\begin{figure*}
\epsfxsize=180 mm
\begin{center}
\leavevmode\epsffile{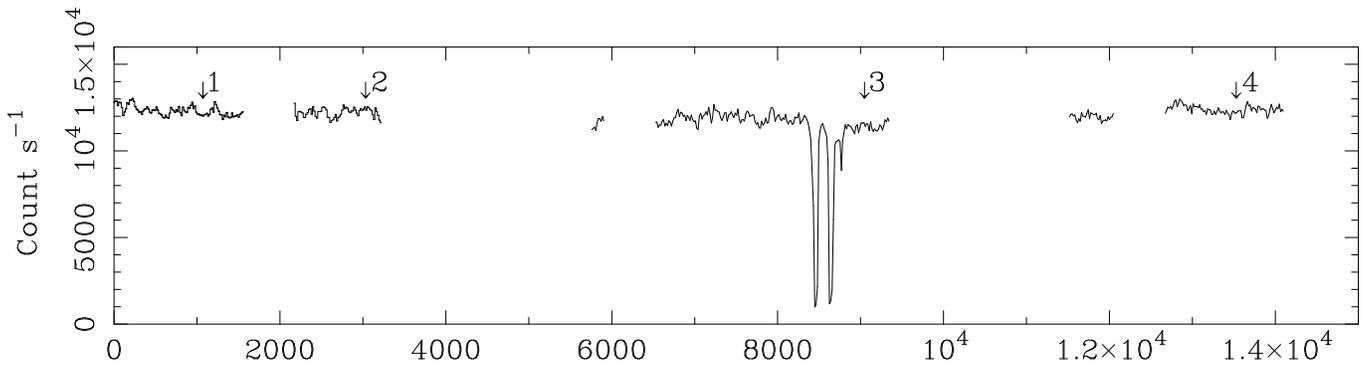}
\end{center}
\caption{PCA light curve in the energy band 2 -- 30 keV with 16 s
binning after deadtime correction \label{fig1}}
\end{figure*}
\section{Analysis and Results}

The observation of \src\ analysed here took place on 1997, Feb 26
lasting 14,600 s and had an exposure of 7.6 ks. 
Data from the PCA instruments in ``Standard 2'' mode are presented.
Data were screened by selecting for angular distance less than
0.02$^{\degmark}$. The PCA consists of 5 Proportional Counting Units
(PCU0 -- 4); spectra were extracted for each PCU
separately but only units 0, 1 and 4 were used because
of the consistently higher values of $\chi^2$ found in fitting data on
the Crab Nebula with PCUs 2 and 3 (Sobczak et al. 1999).
Figure 1 shows the light curve from all 5 PCUs of the PCA with a 
binning of 16 s. Strong dipping is seen at 8.5 ks. 
Four spectra were selected at different times during the observation 
avoiding the dips; these are indicated by arrows 
in Fig.~1, and are labelled spectrum 1 to spectrum 4. The third of these 
(at $\sim$9 ks) follows a dip, and may have slightly reduced intensity. Results 
from 3 further spectra not presented gave similar results.

\begin{figure}
\epsfxsize=80 mm
\begin{center}
\leavevmode\epsffile{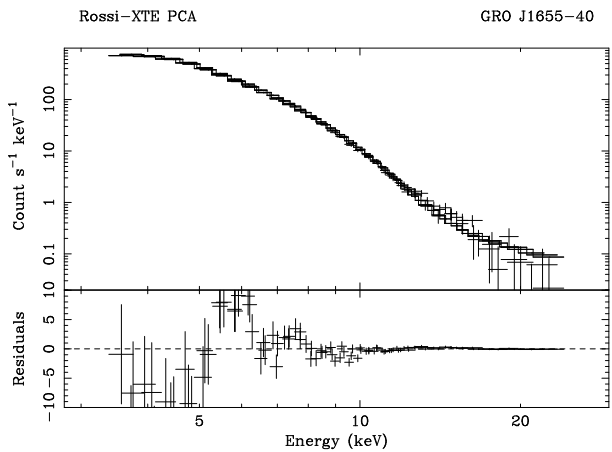}
\end{center}
\caption{Spectrum of a single dataset with residuals; data from PCUs
0, 1 and 4 are shown which were fitted simultaneously \label{fig2}}
\end{figure}

Spectra were accumulated over times averaging 140 s, equivalent to
$\sim$340,000 counts per spectrum. Data were used between 
3.5 -- 25 keV. Primitive channels were regrouped in 2s between channels 30 and 39 
(13 --16 keV) and in 4s above channel 40, and systematic errors 
of 1\% added. Background subtraction was carried out using standard 
background models and instrument responses 
from 1998, Nov 28 used. Spectra from each time interval 
were extracted for PCUs 0, 1 and 4 and these were fitted simultaneously 
with a variable normalization allowed between the PCUs in the fitting. 
A number of spectral models were investigated. Simple one-component models 
were not able to fit the spectra and a good fit to the continuum was obtained
with a two-component model consisting of a disk blackbody plus a power
law component. The luminosity (0.1 -- 100 keV) was $\rm {9.7\times
10^{37}}$ erg s$^{-1}$ with the disk blackbody constituting 89\% of the total and
the power law 11\%. 

However, for the continuum-only model, the $\chi^2$ per degree of
freedom (dof) was
poor, typically 130/91 and positive residuals could be seen in the spectra as shown 
in the example of Fig.~2 (spectrum 3); and these data are re-plotted in the form
of ratios of the data to the best-fit model for each of the 4 datasets
in Fig.~3. Strong line features at $\sim $5.8 keV and $\sim$7.3 keV 
can be seen in all of the spectra. Note that in the ratio plots, the
lower energy part of the line is reduced compared with the higher
energy part because of the decreasing continuum, so that the line
centre appears to be at higher energy than its true value shown in the
residual plots. The 4 spectra were re-fitted with 2 Gaussian lines added to the
model. There was a marked improvement in the quality of fit, with an 
average value of $\chi ^2$/dof of 70/85. Results for all 4 spectra are 
shown in Table 1, where values of $\chi^2$ are compared with values for the continuum 
model alone. F-tests showed that the addition of 2 lines was significant 
at $>>$99.9\% confidence. Fig.~4 shows the spectrum of Fig.~2 with 2 lines 
added to the model. Equivalent widths (EW) were derived for each
Gaussian component, treating the red and blue wings as separate lines
and the red wing had a mean EW of 70 eV and the blue wing a mean EW
of 160 eV. 

Absorption features can also be
seen in the spectra, for example, at $\sim $8.9 keV in spectrum 1
(lowest panel). This feature can be seen in all 4 spectra, and there is some
evidence for small changes in its energy. Spectrum 3 also
apparently has an absorption line or edge at 8.2 keV, and this may be indicative
that the data are not completely out of the dip. To investigate this
point further, dip spectra were examined in relatively shallow dipping
in which the spectrum is not modified in a major way by absorption. A
spectrum was selected in the intensity band 6100--6900 count s$^{-1}$ from
Fig. 1 and the continuum fitted with an appropriate model with partial
covering of the disk blackbody plus power law. In this case, the ratio
plot showed an even stronger depression at $\sim $8 keV than that of
spectrum 3 in Fig.~3,

\begin{figure}
\epsfxsize=80 mm
\begin{center}
\leavevmode\epsffile{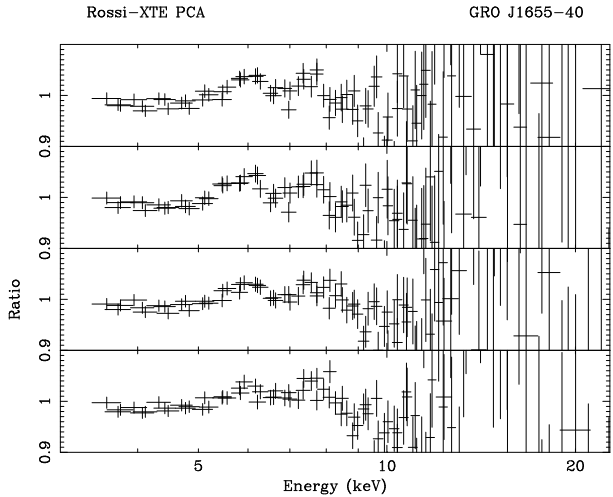}
\end{center}
\caption{Ratios of the data to the models for the 4 spectra indicated
in Fig. 1. The lowest panel shows spectrum 1 \label{fig3}}
\end{figure}

\begin{figure}
\epsfxsize=80 mm
\begin{center}
\leavevmode\epsffile{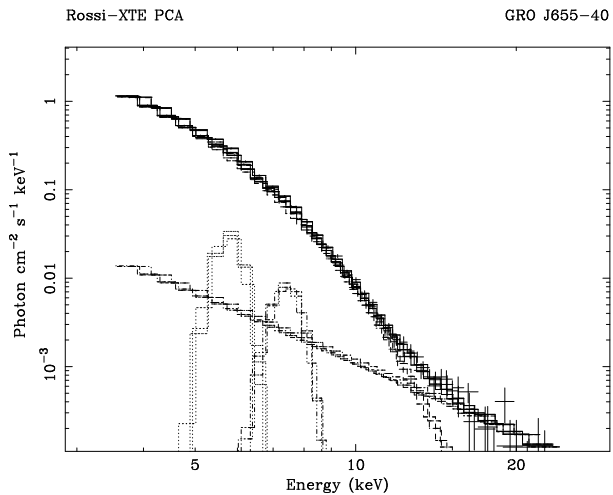}
\end{center}
\caption{Unfolded spectrum of the best-fit to spectrum 3 including
2 Gaussian lines \label{fig4}}
\end{figure}

Although we have detected clear absorption features in the spectra,
we can also ask whether the spectra may be modelled by addition
of a postulated, but undetected, absorption feature at an energy
between the red and blue emission peaks; i.e. by modelling the valley 
between the emission peaks without requiring there to be emission.
The continuum-only model leaves positive residuals at $\sim $5.8 and
$\sim $7.3 keV as can be seen in Fig. 2 and Fig. 3. Since the
continuum is well-fitted (as shown by the $\chi^2$/dof for the best-fit
models), these positive residuals are {\it in no way model dependent},
and so are strong evidence for the disk line. Thus it is {\it impossible}
for the positive residuals to be removed by adding an absorption line.
It is confirmed by spectral fitting tests that the positive residuals
cannot be removed in this way.

\begin{table}
\caption{Line parameters from best fit spectral fitting results for 
spectra 1 -- 4 using a model consisting of
a disk blackbody + power law + 2 Gaussian lines. $\chi^2$ values are
also shown for the best continuum fits without emission lines (RH
column)
\label{1}}
\begin{tabular} {lllrrll}
& $\rm {E_1}$ &$\rm {E_2}$& EW$_1$ &EW$_2$& $\chi^2$/dof & $\chi^2$/dof\\
& (keV)        & (keV) &(eV) &(eV)&\\
\hline
1&$\rm {5.82^{+0.10}_{-0.17}}$&$\rm {7.20^{+0.22}_{-0.59}}$&51  &210  &79/85 &153/91\\
2&$\rm {5.79^{+0.09}_{-0.30}}$&$\rm {7.17^{+0.43}_{-1.93}}$&66  &209 &60/85 &123/91\\
3&$\rm {5.79^{+0.15}_{-0.21}}$&$\rm {7.43^{+0.20}_{-0.58}}$&99  &135 &59/85 &119/91\\
4&$\rm {5.98^{+0.11}_{-0.19}}$&$\rm {7.47^{+0.13}_{-0.22}}$&62  &81 &80/85 &135/91\\
\hline
\end{tabular}
\end{table}

Fitting the spectra with two emission lines was the obvious way of 
investigating the energies of the emission features. The fact that the feature 
energies are consistent with red and blue shifted iron disk line emission
is strong evidence that the features are iron disk line emission 
(the radio jets having ceased to exist) and makes 
it unlikely that these features are coincidentally at these energies and have another 
origin. The intensity of the blue wing of a disk line is however, expected to be higher 
than that of the red wing (Fabian et al. 1989; Laor 1991), whereas we find the blue wing
intensity to be generally rather less than that of the red wing. The absorption features 
detected in {\it ASCA} and in the present work at energies above 6.6 keV can modify the 
observed broad disk line considerably, and we require absorption at about the energy of 
the blue wing for our results to be consistent with a disk line. Consequently, we have 
carried out fitting with a model containg the Laor disk line model in `Xspec' added to 
the disk blackbody plus power law continuum components, plus an absorption line.
Stable, free fitting results were obtained for all four spectra with this model 
and results are shown in Table 2. Values of the restframe energy varied between 6.4 keV
and 6.8 keV with a mean of 6.56 keV. The energy of the absorption line was well-constrained 
since a large residual at a well-defined energy would result from the
blue wing of the Laor model if the absorption line was omitted. Line energies varied between 
7.0 and 7.3 keV. $\chi ^2$/dof values were similar to those obtained with
two emission lines, varying between 60/84 and 87/84. The inner radius of the disk line
emission region r$_1$ was found to be $\sim $10r$_S$; $\rm {r_2}$ was
$\approxgt $50r$_S$, but was poorly constrained. We have 
thus shown that it is possible to fit the spectra with a model based on the assumption 
that the blue disk line wing appears relatively weak because of absorption at 
$\sim $7 keV. It can be argued that the Laor model is preferred because it contains the 
correct line shape and two emission lines do not. On the other hand, the two emission 
line model is better able to determine red and blue energies as there is no complicating 
extra absorption component in the model. However, the line energies must be interpreted 
as an average over the inner disk. Finally, we note that we do not require 
absorption at exactly the energies derived above to reduce the observed flux of the blue 
wing; we are not attempting to fit all of the absorption features in the spectrum and
various absorption features at 7 -- 8 keV would be capable of reducing the blue wing flux. 

\begin{figure}
\epsfxsize=80 mm
\begin{center}
\leavevmode\epsffile{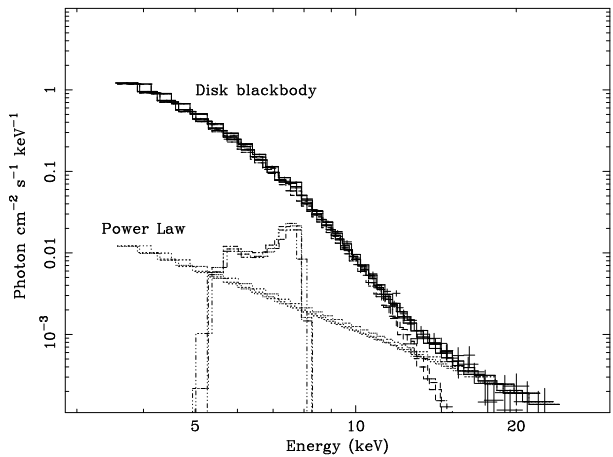}
\end{center}
\caption{Best fit Laor disk line + absorption line model for spectrum 1.
The disk blackbody, power law and Laor components are shown.
The absorption line can not be plotted as it has negative normalization.
The strong dominance of the blackbody emission can be seen.
\label{fig5}}
\end{figure}

Finally, we have tested whether a reflection component can be detected
in GRO\th 1655-40, i.e. a component spectrally different from the incident
power law (see Ross, Fabian \& Young 1999). To do this, we added the
`pexriv' component in Xspec to our best-fit models (Magdziarz \&
Zdziarski 1995), although this model may be inaccurate for high values of $\xi $ (Ross et
al. 1999).  
This was done for the two emission line models and for 
the Laor disk line plus absorption line models. Our results for the disk line energies 
and for the absorption line indicate an ionized medium and so we have allowed the 
ionization parameter $\xi $ in the reflection model to vary between 500 and $\rm {10^4}$. 
The reflection component normalization and power law index were
chained to the values of the power law.
We note however, that without the reflection component, for both the two emission line and 
Laor model fitting, $\chi^2$/dof was already acceptable. Fitting with a reflection
component added showed that there was no reduction in $\chi ^2$. the upper limit flux of 
the pexriv reflection component was 1\% of the total flux at 7 keV. For $\xi $ $<$ $\rm {10^4}$, 
the actual contribution of a reflection component would be less than 1\%. We thus conclude that 
we have not detected a reflection component. If a reflection component exists at a flux level 
of $\approxlt $1\%, it would not be possible for the edge in this component to modify in any 
significant way the values we have derived for red and blue wing energies, widths and EWs.
In this source the blackbody strongly dominates the spectrum with
90\% of the luminosity, see Fig. 5 above, unlike in Cygnus X-1 in the Low State (e.g. Ba\l
uci\'nska-Church et al. 1995) where the power law strongly dominates. The reflection 
component in pexriv is a fraction of the underlying power law of the order of 10\%, and thus we
expect the contribution of a reflection component to be small.

\section{Discussion}

We have presented evidence for a broad emission feature in the X-ray spectrum of \src\
having red and blue wings which we have modelled by two Gaussian lines
and also using the Laor model plus an absorption line. Using the first
model, the line components have high significance as shown by F-tests, and the average
line energies obtained fitting two emission lines from all 4 datasets
analysed are $\rm {5.85\pm 0.08}$ keV and  $\rm {7.32\pm 0.13}$ keV. Given
the inability to fit the spectra without emission lines, and given
these line energies, it is likely that the emission is iron emission.
From the fact that radio emission was switched off before the
observation discussed, the lines almost certainly originate in the disk.  

We have also fitted the spectra with a model consisting of continuum
components plus a Laor disk line plus an absorption line, and conclude
that the blue wing intensity less than expected can be explained by
absorption at $\sim $7 keV reducing the observed flux of the blue wing.
This can be either iron $\rm {K_{\alpha}}$ or $\rm {K_\beta}$
absorption. The results for this model give somewhat smaller restframe energies
than derived from the two emission line model, the mean value being
6.56$\pm $0.14 keV compared with 6.88$\pm $ keV.

Firstly, we will compare our results with those obtained using {\it ASCA}.
In the observations of 1994 -- 1996, the energies at which iron absorption 
lines were detected were $\sim
$6.6 keV and $\sim $7.7 keV for the source at lower intensity, and
$\sim$7 keV and $\sim $8 keV when brighter (Ueda et al 1998). In the {\it RXTE} data
however, we see evidence for an absorption feature at $\sim $9 keV and for
an absorption feature at 8 keV in spectrum 3. In the {\it ASCA} observation made on
1997, Feb 26--28 which included the {\it RXTE} observation, the total
spectrum containing all data showed broad absorption at 6.7 keV thought
to be a blend of He-like and H-like iron $\rm {K_{\alpha}}$ lines
(Yamaoka et al. 1999). This energy corresponds approximately to the
position of the neck between the red and blue peaks that we detect.
The {\it ASCA} spectra do show some evidence for emission however (i.e
positive residuals), particularly when the source is not very bright (Fig. 3 of Ueda et al.
1998). It is not clear at this stage why emission features were not seen
more clearly in {\it ASCA}; one possibility is that the emission
may vary with time, and the {\it ASCA} spectra were integrated over
relatively long periods leading to smearing of the emission.

\begin{table}
\caption{Best fit spectral fitting results for spectra 1 -- 4 using a
model consisting of
a disk blackbody + power law + Laor disk line model + absorption line. 
\label{2}}
\begin{tabular} {lllrrll}
& $\rm {E_{rest}}$ &$\rm {r_1}$& $\rm {r_2}$ &$\rm {E_{abs}}$& $\chi^2$/dof \\
& (keV)        &($\rm {r_S}$)       &($\rm {r_S}$)   &(keV)  &\\
\hline
1&$\rm {6.79^{+0.13}_{-0.33}}$ &$\rm {10^{+24}_{-6}}$ &$\rm {41^{+115}_{-14}}$&$\rm {7.30^{+0.31}_{-0.67}}$  &87/84\\
2&$\rm {6.54^{+0.08}_{-0.16}}$&$\rm {13^{+10}_{-12}}$  &$\rm {56^{+140}_{-24}}$&$\rm {7.10^{+0.15}_{-0.19}}$  &75/84\\
3&$\rm {6.43^{+0.06}_{-0.05}}$&$\rm {9^{+33}_{-9}}$  &$\rm {75^{+125}_{-45}}$&$\rm {6.99^{+0.06}_{-0.21}}$  &60/84\\
4&$\rm {6.47^{+0.21}_{-0.16}}$&$\rm {9^{+12}_{-4}}$   &$\rm {40^{+160}_{-15}}$&$\rm {6.96^{+0.06}_{-0.17}}$  &80/84\\
\hline
\end{tabular}
\end{table}

Using our results from fitting two emission lines,
the relativistic Doppler formula can be used with the mean energies of the red and blue
wings E$_1$ = 5.85 keV and E$_2$ = 7.32 keV to solve for the velocity $\beta $ = v/c 
and the restframe energy of the disk line. This gives a mean restframe energy of 
$\rm {6.88\pm 0.12}$ keV and a mean $\beta $ = $\rm {0.33\pm 0.02}$ for an inclination 
angle of $\rm {70\degmark}$. The mean of $\rm {E_1}$ and $\rm {E_2}$ has an average of 
6.58$\pm 0.10$ keV over the four spectra so that the redshift is 0.3
keV, i.e. z = 0.046. This is partly gravitational redshift and partly the transverse Doppler effect. 
A gravitational redshift of 0.046 would be produced by emission at $\sim $12r$_S$. 

Results from the Laor model give a mean restframe energy of 6.56$\pm
0.14$ keV. We conclude that the restframe energy is between 6.4 and 7.0 keV
indicating Fe $\rm {K_{\alpha}}$ emission. The exact ionization state
is not clear; the mean of 6.56 keV implies Fe XXII. 
The Laor model fitting further provided the inner
and outer radii of the disk line emission region r$_1$ $\sim $10r$_S$
and r$_2$ $\approxgt $50r$_S$. It may be thought that
emission from different radii would tend to smear out the wings;
however, simulations show that this does not take place for emission
between 10 -- 100r$\rm {_S}$, or even for emission between
10 -- 200r$\rm {_S}$. The inner radius of the emission region is more
important, and if we allow emission from 1 -- 10r$\rm{_S}$, there is
smearing out because of the strong emission from inner radii and changing 
energy shifts. However, it is likely that the disk inside 10r$_S$ is
totally ionized by X-ray emission so that no lines are produced and
large-scale smearing out does not occur.

The {\it ASCA} observation of 1997, February spanning one orbital cycle, detected 
an absorption line at $\sim $6.8 keV at all orbital phases, showing that
the absorbing material was not confined to part of the accretion disk structure 
(Yamaoka et al. 1999) as in LMXB. Moreover, the line energies 
seen generally in {\it ASCA} were H-like or
He-like showing that the absorbing plasma was highly ionized.
Our observation of highly broadened iron emission clearly shows the
line to originate in the inner, highly ionized, disk. The location
of the absorber is however, not so clear.

Further observation and detailed analysis is clearly needed to explain
the observed spectral features which are complex, the absorption
features in particular, appearing to change with source intensity.
GRO\th J1655-40 offers probably the best  opportunity of 
studying disk lines strongly affected by gravitational and Doppler 
effects because of the high inclination at which it is seen. 
Our detection of the red and blue shifted wings at energies of 5.8 and
7.3 keV
is direct evidence for the black hole, since a splitting as wide as this
cannot be produced by a neutron star and is the first detection
of a red and blue shifted disk line in a Galactic source.

\section{Acknowledgments}

We would like to thank Prof. Hajime Inoue for valuable discussions on
the relation between the emission features reported here and absorption
features discovered with {\it ASCA}.


\begin{thebibliography}{}

\bibitem[]{}
Ba\l uci\'nska-Church M., Church M. J., Charles P. A.,
Nagase F., LaSala J., Barnard R., 1999, MNRAS {\it In Press}
 
\bibitem[]{}
Ba\l uci\'nska-Church M., Belloni T., Church M. J., Hasinger G.,
A\&A 302, L5

\bibitem[]{}
Fabian A. C., Rees M., Stella L., White N. E., 1989, MNRAS, 238, 729

\bibitem[]{}
Hjellming R. M., Rupen M. P., 1995, Nature, 375, 464

\bibitem[]{}
Hunstead R. W., Campbell-Wilson D., 1996, IAU Circ. 6410

\bibitem[]{}
Laor A., 1991, ApJ, 376, 90

\bibitem[]{}
Magdziarz P., Zdziarski A. A., 1995, MNRAS, 273, 837

\bibitem[]{}
Mirabel I. F., Rodriguez L. F., 1994, Nature, 371, 46

\bibitem[]{}
Orosz J. A., Bailyn C. D., 1997, ApJ, 477, 876

\bibitem[]{}
Ross R. R., Fabian A. C., Young A. J., 1999, MNRAS 306, 461

\bibitem[]{}
Sobczak G. J., McClintock J. E., Remillard R. A., Bailyn C. D., Orosz J.
A., 1999, ApJ, 520, 776

\bibitem[]{}
Thorne K. S., 1974, ApJ, 191, 507

\bibitem[]{}
Tingay S. J., et al. 1995, Nature, 374, 141

\bibitem[]{}
Tingay S. K., 1999 {\it Priv. Comm.}

\bibitem[]{}
Ueda  Y., Inoue H., Tanaka K., Ebisawa K., Nagase F., Kotani T.,
Gehrels N., 1998, ApJ, 492, 782

\bibitem[]{}
Yamaoka K. et al. 1999, {\it In Preparation}

\bibitem[]{}
Zhang S. N., Wilson C. A., Harmon B. A., Fishman G. J., Wilson R. B.,
Paciesas W. S., Scott M.,  Rubin B. C., 1994, IAU Circ., 6046

\bibitem[]{}
Zhang S. N., Harmon B. A., Paciesas W. S., Fishman G. J., 1995,
IAU Circ., 6209

\end{thebibliography}
\end{document}